\documentclass[preprint]{revtex4}
\usepackage{amsmath,amssymb,amsfonts}
\usepackage{bm}
\usepackage{graphicx}
\usepackage{times}
\usepackage{color}
\usepackage[colorlinks,bookmarks=false,citecolor=blue,linkcolor=red,urlcolor=blue]{hyperref}

\newcommand{\fref}[1]{Fig.~\ref{#1}}
\newcommand{\eref}[1]{Eq.~(\ref{#1})}

\newcommand{\normwidth}{0.8\columnwidth}

\newcommand{\miniwidth}{0.49\columnwidth}

\begin{document}

\title{Phase diagram of Holstein-Kondo lattice model at half-filling}
\author{Reza Nourafkan}
\affiliation{Department of Physics, Sharif University of Technology,
P.O.Box: 11155-9161, Tehran, Iran}
\author{Nasser Nafari}
\affiliation{Institute for Studies in Theoretical Physics and
Mathematics, P.O.Box: 19395-5531, Tehran, Iran}

\begin{abstract}
We study the Kondo lattice model which is modified by the Holstein
term, involving both the Kondo exchange coupling and the
electron-phonon coupling constants, characterized by $J$ and $g$,
respectively. The model is solved by employing the dynamical
mean-field theory in conjunction with exact diagonalization
technique. A zero temperature phase diagram of symmetry unbroken
states at half filling is mapped out which exhibits an interplay
between the two interactions and accounts for both spin and charge
fluctuations. When the Kondo exchange coupling is dominant the
system is in Kondo insulator state. Increasing $g$ for small values
of $J$ leads to a Kondo insulator-metal transition. Upon further
enhancement of $g$ a transition to the bipolaronic insulating phase
takes place. Also a small region with non-Fermi liquid behavior is
found near the Kondo insulator-metal transition.
\end{abstract}
\maketitle There has been a continued interest in a class of
compounds called heavy fermion semiconductors, which exhibit a spin
and a charge gap at low temperatures typically ranging between $1$
and $100$ meV \cite{Riseborough,Misra}. In contrast to the ordinary
band insulators, these two gaps are different, indicating a
separation of the spin and charge degrees of freedom brought about
by correlation effects. The gap formation in heavy fermion
semiconductors is attributed to the renormalized hybridization
between a broad band of conduction electrons and a nearly flat band
of strongly correlated $f$-electrons.

The Kondo lattice model (KLM) at half-filing is considered to be a
good starting point for investigating the properties of the heavy
fermion semiconductors. In this model, at each lattice site a local
moment interacts with the spin of a conduction electron, and thus,
results in complex correlation effects between them. In fact, a
conduction and a localized electron with antiparallel spins undergo
a spin-flip process, causing itinerant electrons to leave a trace of
their spin exchange at each localized spin site. As a result, the
direction of a localized spin is affected by the history of the
electrons passing through it. There are similar correlation effects
in the periodic Anderson model due to the dynamical aspects of the
localized electrons.




Experiments involving the Kondo insulators at high magnetic fields
indicate the closure of the Kondo insulating gap, exemplifying a
transition from the Kondo insulator to a correlated metal
\cite{Jaime,Cooley}. It is expected that the electron-phonon (e-ph)
interaction leads to similar results. Many experiments suggest that
the e-ph effects are important in describing a number of
observations such as the existence of an unusual phonon softening in
the Kondo lattice, CeCu$_2$, which is indicative of coupling between
electrons and phonons \cite{Loewenhaupt}. Furthermore, the lattice
plays an important role in some heavy fermion compounds, called
14-1-11, where various properties can be altered through
isoelectronic substitutions \cite{Burch}. In fact, it is believed
that the coupling between phonon modes and the Kondo effect could
manifest new material properties, such as non-Fermi liquid behavior
and unconventional superconductivity \cite{Yotsuhashi,Hotta,Nayak}.

Even less studies has been devoted to the role played by lattice
vibrations in these compounds. The role of the lattice vibrations is
not trivial, but if, on general grounds, the minimal effect of e-ph
coupling is a phonon-retarded attraction between conduction
electrons with opposite spins, then the spin excitation has a gap
while the charge excitation, depending on the strength of the e-ph
coupling, can be either gap-full or gapless. Therefore, there arises
a competition between the spin- and the charge-fluctuations whose
behavior is determined, on the one hand, by the relative strength of
the Kondo exchange between the conduction electrons and the
localized moments and, on the other hand, by the conduction
electron-phonon coupling leading to a complicated phase diagram.


It is the goal of this paper to investigate the dynamical
competition between the e-ph and Kondo interactions. A natural way
of incorporating the e-ph coupling in the KLM is to add the Holstein
coupling term to its hamiltonian. In the Holstein coupling the
phonon variables are coupled to the local density of the conduction
electrons. In this paper, we will present the zero temperature phase
diagram of the Holstein-Kondo lattice model (H-KLM) at half-filling.
The focus is on the transition between the unbroken symmetry ground
state as the e-ph and Kondo interactions parameters, $J$ and $g$,
are varied.

The H-KLM Hamiltonian is defined by:
\begin{eqnarray}
\label{eq:Hol_Kon}
H=&-&t\sum_{\langle i,j \rangle \sigma}{\left( c^\dagger_{i\sigma}c_{j\sigma} + c.c. \right)} +
  \frac{J}{2}\sum_{i,\alpha \beta} {\bm{S}_{i}.\left( c^\dagger_{i\sigma} \bm{\sigma}_{\alpha \beta}c_{i\sigma}    \right)} \cr
  &+& g\sum_{i}{\left( n_{i} - 1\right)\left( b^\dagger_{i} + b_{i} \right)} +
  \Omega_{0}\sum_{i} b^\dagger_{i}b_{i},
\end{eqnarray}
where $c_{i\sigma}\left(c^\dagger_{i\sigma} \right)$  and
$b_{i}\left(b^\dagger_{i} \right)$ are, respectively, destruction
(creation) operators for itinerant electrons with spin $\sigma$ and
local vibrons of frequency $\Omega_{0}$ on site $i$, $n_{i}$ is the
electron density on site $i$, $\bm{S}_{i}$ is the spin operator for
the localized spin on site $i$, $\bm{\sigma}$ is a pseudo-vector
represented by Pauli spin matrices, $t$ stands for the itinerant
electrons hopping matrix elements between the nearest-neighbor
sites, $J$ is the coupling strength between itinerant electrons and
localized spins, and $g$ denote the  electron-phonon coupling. We do
not consider the coulomb repulsion term between itinerant electrons,
because it tends to suppress the double occupation of sites and in
our model the exchange coupling, $J$, already does the same thing.


Our calculations are based on the dynamical mean field theory
\cite{Georges}, a powerful, non-perturbative tool to study the
properties of strongly correlated systems, which allows us to treat,
on equal footing, the two kinds of interactions present in our
model. This technique, which becomes exact in the limit of infinite
coordination number, reduces the full lattice many-body problem to a
local impurity embedded in a self-consistent effective bath of free
electrons, mimicing the effect of the full lattice on the local
site. A self consistency condition links the effective impurity
model to the original lattice problem. Adopting a semi-circular
density of states (DOS) $\rho_{0}(\epsilon)=(2/\pi
D)\sqrt{D^{2}-\epsilon^{2}}$ of the noninteracting system,
corresponding to a Bethe lattice with the half bandwidth $D$, the
self-consistency relation imposed on the DMFT solution is given by
\begin{equation}\label{eq:SC}
\frac{D^{2}}{4}G(i\omega_{n})=\sum_{k}{\frac{V_{k}^{2}}{i\omega_{n}-\epsilon_{k}}},
\end{equation}
where $\epsilon_{k}$ and $V_{k}$ are the energies and the
hybridization parameters of the effective impurity model (bath
parameters). We use exact diagonalization (ED) technique to solve
the effective impurity model \cite{Caffarel}. This solver allows us
to access the ground state properties of the system with a finite
energy resolution. The ED technique consists of restricting the sum
in \eref{eq:SC} to a small number of levels, and moreover, it
truncates the infinite phonon Hilbert space. The ground state and
the Green's function of our discretized model are determined via the
Lanczos procedure and the self-consistency equation in turn allows
us to derive a new set of bath parameters. The process is iterated
until convergence is reached. In the theory of Mott transition, the
investigation of the paramagnetic (PM) phase has been very fruitful
providing a lucid understanding of the finite temperature state,
above the magnetic order in many compounds. We pursue a similar
approach in our investigation and study the PM state. We force the
system to be in a paramagnetic state by averaging the spin up and
spin down to study the underlying normal state. In all our
calculations presented here the convergence of truncation has been
checked.

\begin{figure}
  \begin{center}
    \includegraphics[width=\normwidth]{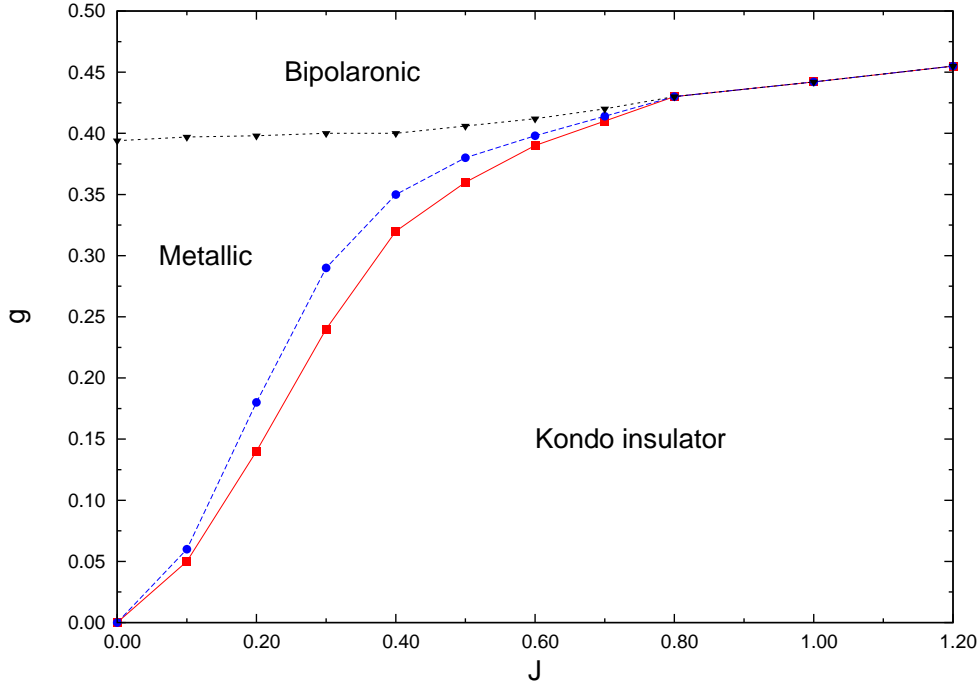}
    \caption{Zero temperature phase diagram of the unbroken symmetry Holstein-Kondo lattice model at  half-filling. The model shows three different phases: metallic, bipolaronic and Kondo insulating phase. A narrow region with non-Fermi liquid character is seen near the Kondo insulator-metal transition. } \label{fig:phase diagram}
  \end{center}
\end{figure}

\fref{fig:phase diagram} shows the $T=0$ phase diagram of the
half-filled H-KLM in the parameter space of $J$ and $g$ with
$D=2t=2$ and $\Omega_{0}/t=0.2$. All types of long-range order are
excluded. Three different phases are distinguished: metallic phase
and the bipolaronic and Kondo insulating phases. In what follows, a
detailed discussion of the phase diagram of these systems will be
presented. The Kondo lattice model ($g=0$) and Holstein model
($J=0$), which are special limiting cases of the H-KLM, have been
extensively studied using the DMFT. The ground state of KLM is the
Kondo insulating phase with a spin and a charge gap for all $J$
values \cite{Costi}. For the Holstein model, the ground state is
metallic. The metallic phase is found to be a Fermi liquid, in the
sense that, the Luttinger sum rule $\rho(0)=\rho_{0}(0)$ for the
spectral function $\rho(\omega)=-ImG(\omega+i0^{+})/\pi$, or equally
stated, the limit of $ImG(i\omega_{n})\rightarrow -1$ as
$\omega_{n}\rightarrow 0$, is satisfied ($\omega_{n}$ is the
Matsubara frequency). Upon increasing $g$, the conduction electrons
lose their mobility, eventually acquiring polaronic character, in
which the presence of an electron is associated with a finite
lattice distortion. Also, the same e-ph coupling can cause any two
polarons to attract and form a bound pair in real space, called
bipolaron \cite{Capone1}. In the absence of pair hopping, the
bipolaron formation would cause the system to undergo a first order
metal to bipolaronic insulating phase transition at the  critical
coupling $g_{c}$ \cite{Koller1,Jeon}. Meyer et al. have reported
that there is a coexistence region near $g_{c}$, which is reduced as
the phonon frequency $\Omega_{0}$ is decreased and disappears for
$\Omega_{0} .leq. 0.10 D$ \cite{Meyer}. The bipolaron formation may
be accommodated by reconstructuring the system into a phase
separated state \cite{Capone2} or a charge ordered state in which
the doubly occupied and empty sites alternate in real space
\cite{Pietig}.

\begin{figure}
  \begin{center}
    \includegraphics[width=\miniwidth]{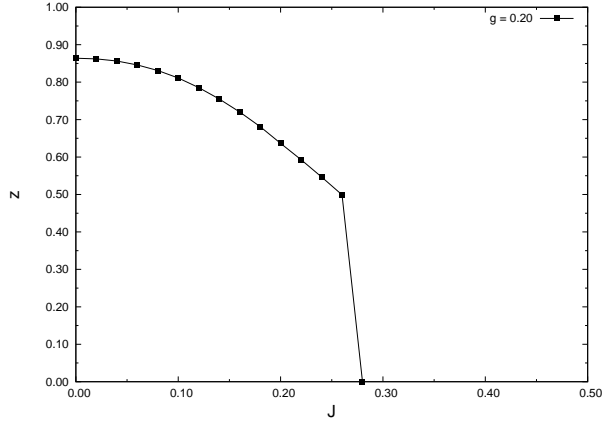}
    \caption{Behavior of quazi-particle weight for different values of $g$.} \label{fig:z}
  \end{center}
\end{figure}

At small fixed $J$-values, with increasing e-ph coupling, a
continuous transition to a metallic state occurs at a critical
coupling $g_{1c}(J)$, whose value increases with increasing $J$.
This behavior is physically expected. An increase in $J$ leads to a
larger insulating gap, and this in turn, leads to the suppression of
the charge fluctuations which would otherwise couple to phonons. As
a result a transition to metallic state occurs at larger e-ph
coupling. We Also find that the metallic phase near $g_{1c}$ shows
non-Fermi liquid character. Further increase of $g$ causes a
metal-bipolaronic phase transition taking place at a critical
coupling $g_{2c}$. As it can be distinguished, a Holstein coupling
is weakly affected by exchange coupling between conduction electrons
with local spins. The metallic state becomes more correlated as $g$
or $J$ is increased. This is reflected in the decreasing behavior of
the quasiparticle weight $z=1/[1-Im\Sigma(i\omega_{0})/\omega_{0}]$
when $g$ or $J$ is increased (\fref{fig:z}).

\begin{figure}
  \begin{center}
    \center{\includegraphics[width=\normwidth]{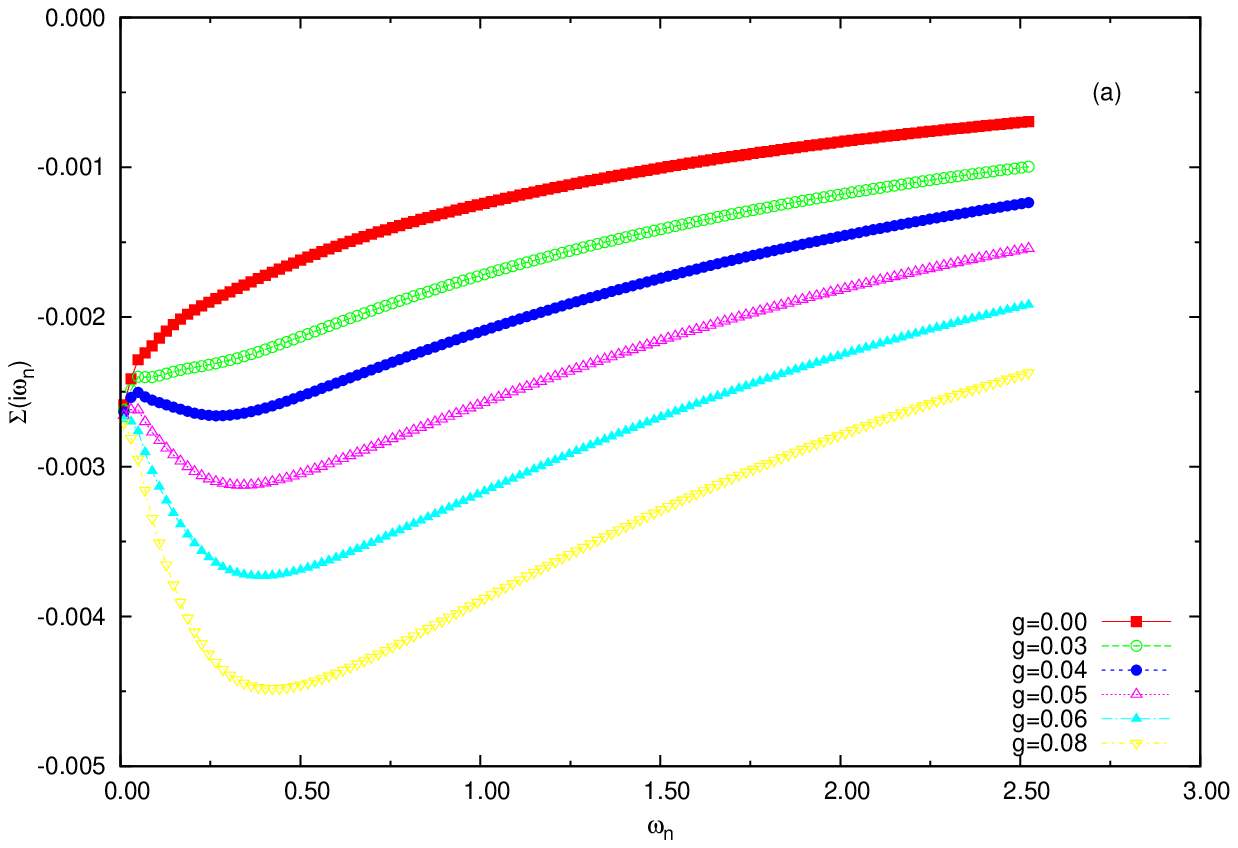}}
    \center{\includegraphics[width=\normwidth]{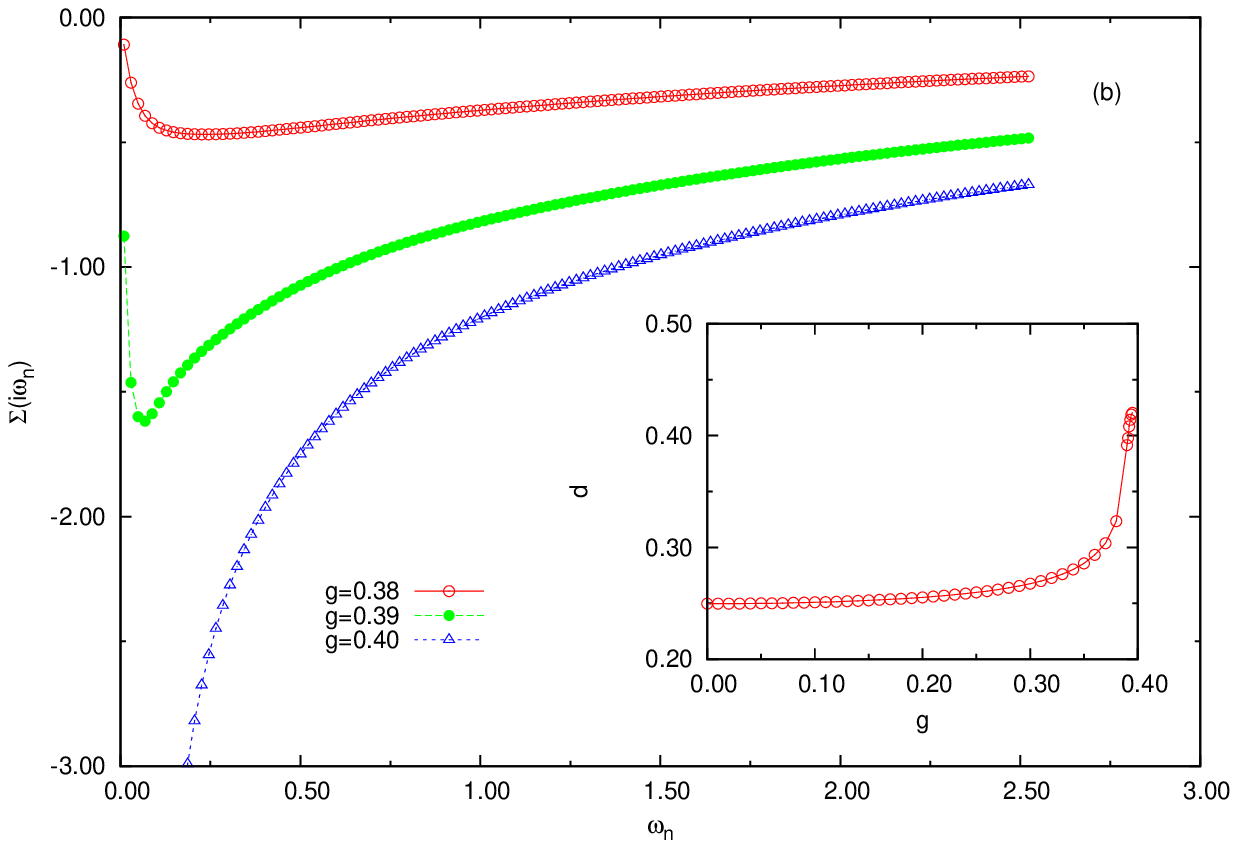}}
    \caption{Imaginary part of electron self-energy, $Im\Sigma (i\omega_{n})$, obtained  at different values of $g$ in the vicinity of both phase transitions, with fixed $J=0.1$. Panel (a): $Im \Sigma (i\omega_{n})$ in the vicinity of the transition from the Kondo insulator state to the metallic state. Changing the $Im\Sigma (i\omega_{n})$ behavior as $\omega_{n}\rightarrow 0$ from diverging to extrapolating to zero shows the insulator-matallic phase transition. Panel (b): $Im \Sigma (i\omega_{n})$ in the vicinity of the transition from the metallic state to bipolaronic state. Inset: the double occupancy $d=<n_{\uparrow}n_{\downarrow}>$ as a function of $g$. The transition from the metallic state to bipolaronic state is clearly visible by observing when the double occupancy's jump to $\approx1/2$ begins to set in.}
 \label{fig:self-energies}
  \end{center}
\end{figure}

\fref{fig:self-energies} shows the imaginary part of electron
self-energy, $Im\Sigma(i\omega_{n})$, for $J=0.1$ and several values
of $g$ in the vicinity of both phase transitions. For small $g$
values, the imaginary part of the self-energies diverge as
$\omega_{n}\rightarrow 0$, indicating the presence of a charge gap
(See panel a). Increasing $g$ causes the system to change its phase
from an insulator to a bad metal in the sense that its self-energy
extrapolates to a finite value $Im\Sigma(i0^{+})\equiv \Gamma(J)
\neq 0$ for $g\geq g_{1c}$. Hence, a finite lifetime is found at the
Fermi level for a narrow range of e-ph couplings near the $g_{1c}$,
indicating that well defined quasiparticles do not exist in this
range. The violation of the Luttinger sum rule in this region is
also seen from $ImG(i\omega_{n})$, which tends to a negative
constant $c<0$ in the limit of $\omega_{n}\rightarrow 0$, with $c<
\pi \rho_{0}(0) = 1$. Although the discretness of the spectra
obtained in the exact diagonalization techniqe does not allow us to
unambiguously identify the non-Fermi liquid region, we believe that
the spectral function at $g=0$ displays a narrow insulating gap,
whose width is proportional to the value of $J$, with two peaks on
each side.  For a fixed $J$, increasing $g$ causes the low-energy
spectrum widen and are also suppressed. If these peaks overlap
before being damped completely, a narrow pseudogap forms near the
Fermi level, $E_{F}$. With further enhancement of $g$, there is a
rapid shallowing of the pseudogap till finally a quasiparticle peak
forms at $E_{F}$. At this stage, the system will have a Fermi-liquid
character. Upon increasing $g$ further, there is a weakly narrowing
of quasiparticle peak until it disappears at the second critical
value of e-ph coupling $g_{2c}$ where a gap opens. A more detailed
results on the spectra might be obtained by the numerical
renormalization group technique. The inset of panel (b) shows double
occupancy $d=<n_{\uparrow}n_{\downarrow}>$ as a function of $g$.
There is no signature of the Kondo insulator-metal transition in the
double occupancy, but at $g_{2c}$ the double occupancy jumps
suddenly to $d\approx1/2$, indicating a discontinuous transition to
bipolaronic phase.

\begin{figure}
  \begin{center}
    \center{\includegraphics[width=\normwidth]{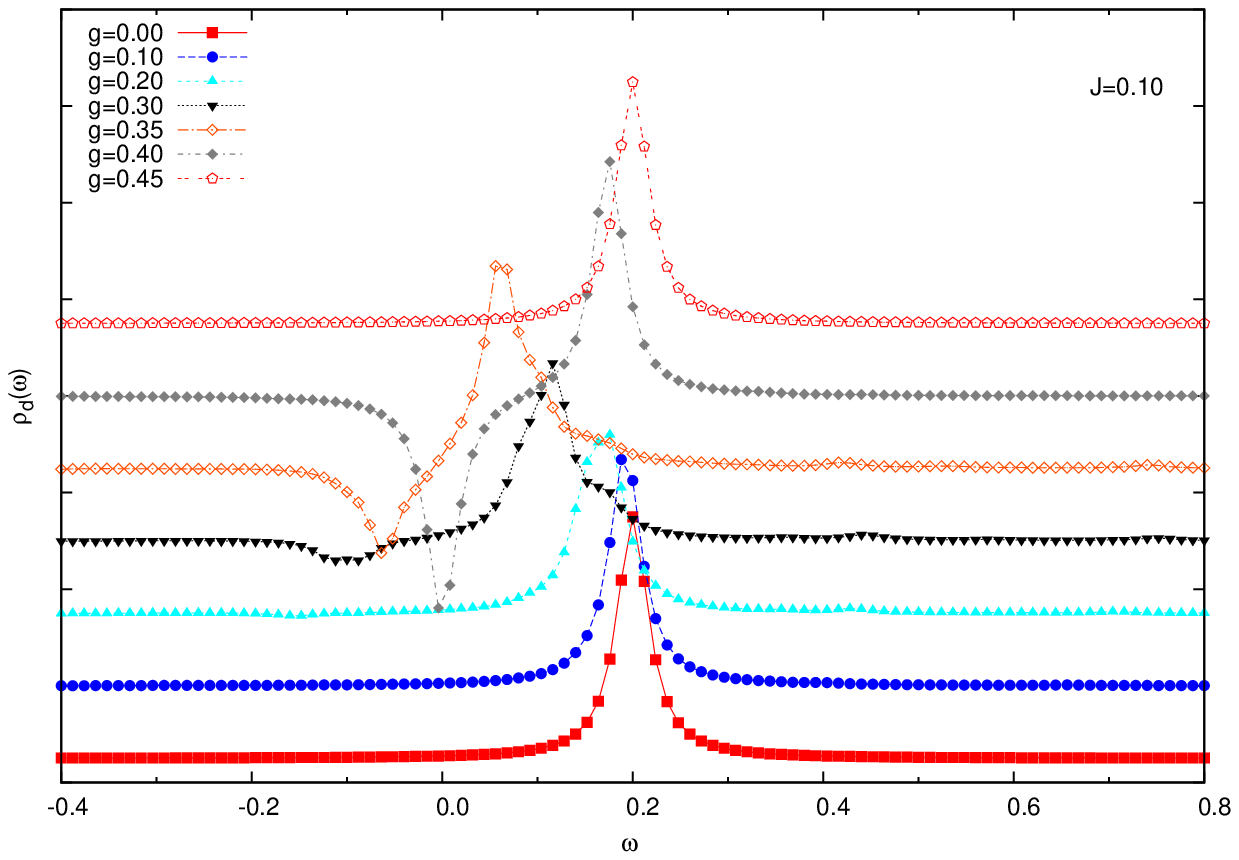}}
    \center{\includegraphics[width=\normwidth]{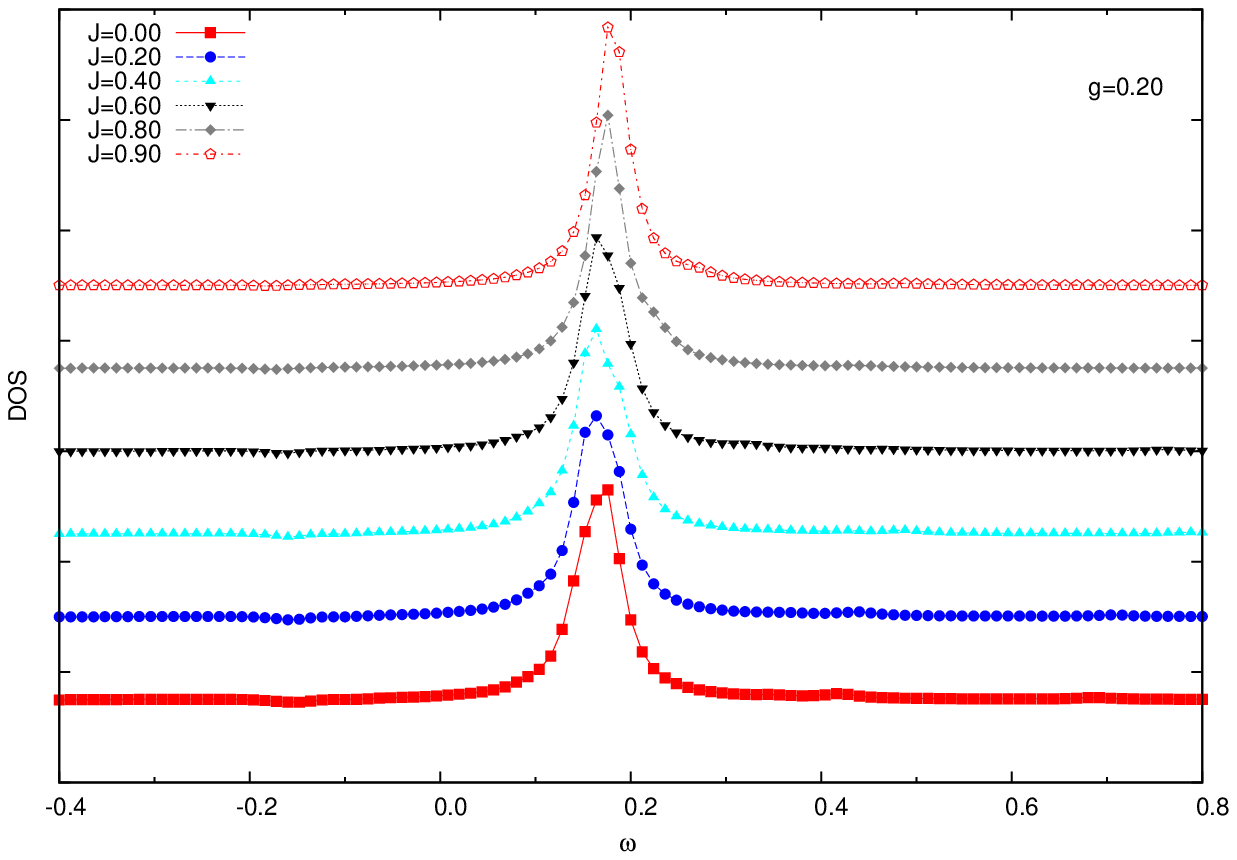}}
    \caption{Phonon spectral function for different values of $g$. The bare phonon frequency is $\Omega_{0}=0.2$ and a Lorentzian broadening with the full width at half maximum of $0.02$ has been implemented. Panel (a): Spectral function for $J=0.1$ and various values of $g$. A cosiderable phonon softening is seen upon approaching the transition to the bipolaronic insulator. Panel (b): Spectral function for $g=0.2$ and various values of $J$. The transition to Kondo insulator does not obviously affect the phonon spectral function.} \label{fig:phonon-spectral}
  \end{center}
\end{figure}

\fref{fig:phonon-spectral}(a) shows the phonon spectral function,
$\rho_{ph}(\omega)=-Im d(\omega+i0^{+})/\pi$, for $J=0.1$ as a
function of e-ph interaction strengths. The phonon Green's function
is defined by $d(\omega)=\ll b_{i};b^{\dagger}_{i}\gg_{\omega}$. The
figure illustrates how the phonon mode is softened with increasing
$g$. The softening phonon mode is a manifestation of a lattice
instability as in structural phase transitions. A stability is
restored by the condensation of the unstable mode. It results in a
nonzero expectation value of the phonon operator ($<b>\neq0$) or in
large average number of excited phonons in the ground state. The
appearance of negative spectral function for $\omega <0$, when the
bipolaronic state is approached implies that there is a large
increase in the lattice displacement. In the bipolaronic state, the
phonon mode hardens back to the bare mode as $g$ assumes values
greater than $g_{c2}$. This is due to the fact that screening is not
effective in an insulating state. This is the  same behavior which
had already been seen for pure Holstein model \cite{Koller2}. Panel
(b) of \fref{fig:phonon-spectral} shows the phonon spectral function
for $g=0.2$ and various values of $J$. The phonon mode gradually
hardens back to $\Omega_{0}$, as the $J$-values increase. We observe
no signature of a transition to Kondo insulator in the phonon
spectrum. The effect of increasing $J$ is to suppress contiuously
the charge fluctuations which results in a decoupling of electrons
and phonons causing the phonon peak to exhibit hardening. In
contrast to the Holstein-Hubbard model results, where softening is
absent in Mott insulator phase and phonons are effectively decoupled
from electrons \cite{Koller2}, here the hardening of the phonon peak
takes place very slowly.

In concluson, we have studied the Holstein-Kondo lattice model at
half-filling. We find that the model presents the physics of the
Kondo insulator when the exchange coupling, $J$, plays a dominant
role and a transition to correlated metal takes place for small $J$
and intermediate e-ph coupling, $g$. Moreover, a bipolaronic-metal
insulator takes place for small $J$ and large $g$. We also find a
small region with non-Fermi liquid character near the Kondo
insulator-metal transition. The remaining interesting questions will
be how the phase diagram and nature of transitions will change as
$\Omega_{0}$ or electron density is changed. It is also interesting
to study the symmetry breaking states such as the antiferromagnetic
and superconducting states. Works in this direction are in progress
and will be reported in a separate publication.

\begin{acknowledgements}
\end{acknowledgements}

\bibliographystyle{prsty}

\end{document}